\documentclass[a4paper, final, 12pt]{article}
\usepackage{amsmath, amssymb, latexsym, amscd, amsthm,amsfonts,amstext}
\usepackage[mathscr]{eucal}
\usepackage{graphicx}
\usepackage{subfig}
\usepackage{float}
\usepackage{color}
\usepackage{hyperref}
\usepackage[utf8]{inputenc}
\usepackage[english]{babel}
\usepackage{bbm}

\newtheorem{theorem}{Theorem}[section]

\newtheorem{definition}{Definition}[section]

\numberwithin{equation}{section}
\input{tcilatex}
\begin{document}

\title{Lipschitz Stability Estimate and Uniqueness in the Retrospective
Analysis for The Mean Field Games System via Two Carleman Estimates }
\author{ Michael V. Klibanov \thanks{
Department of Mathematics and Statistics, University of North Carolina at
Charlotte, Charlotte, NC, 28223, USA, mklibanv@uncc.edu} \and Yurii
Averboukh \thanks{%
Krasovskii Institute of Mathematics and Mechanics, Yekaterinburg, S.
Kovalevskaja st., 16, 620990, Russia, ayv@imm.uran.ru} }
\date{}
\maketitle

\begin{abstract}
A retrospective analysis process for the mean field games system (MFGS) is
considered. For the first time, Carleman estimates are applied to the
analysis of the MFGS. Two new Carleman estimates are derived. They allow to
obtain the Lipschitz stability estimate with respect to the possible error
in the input initial and terminal data for a retrospective problem for MFGS.
This stability estimate, in turn implies uniqueness theorem for the problem
under the consideration. The idea of using Carleman estimates to obtain
stability and uniqueness results came from the field of Ill-Posed and
Inverse Problems.
\end{abstract}

\textbf{Key Words}: mean field games system, retrospective analysis, two
Carleman estimates, Lipschitz stability estimate, uniqueness

\textbf{2020 MSC codes}: 91A16, 35R30

\section{Introduction}

\label{sec:1}

This paper is concerned with the retrospective analysis of the second order
mean field games system (MFGS) in a bounded domain. We use an
overdetermination in the terminal condition. For the first time, the
Lipschitz stability estimate for this problem with respect to the possible
noise in the initial and terminal data is proven here. This estimate implies
uniqueness. The authors are unaware about published results of this nature
for the MFGS.

The methodology of this work came from the field of Ill-Posed and Inverse
Problems, in which the first author has been working for many years, see,
e.g. \cite{BukhKlib,BK,Klib84,Klib92,KT,Ksurvey,KAPNUM,KL}. In fact, the
problem considered here is close to the classical ill-posed Cauchy problems
for PDEs, in which case an overdetermination in a boundary condition takes
place, see, e.g. \cite{KAPNUM}, \cite[Chapter 3]{Isakov}, \cite[Chapter 2]%
{KT}, \cite[Chapters 2,5]{KL}, \cite[Chapter 4]{LRS}. \textbf{\ }

Our technique relies on two new Carleman estimates. To our best knowledge,
this is the first time when Carleman estimates are applied in the mean field
games theory. Carleman estimates were first introduced in the seminal work
of Carleman of 1939 \cite{Carl}. Since then they have been used for proofs
of stability and uniqueness results for ill-posed Cauchy problems for PDEs,
see, e.g. \cite{H,Isakov,KT,KL,LRS}. In 1981 Bukhgeim and Klibanov \cite%
{BukhKlib} have introduced the tool of Carleman estimates in the field of
coefficient inverse problems (CIPs) for proofs of global uniqueness theorems
and stability estimates for these problems. There are many publications,
which use the framework of \cite{BukhKlib} for proofs of such results. Since
the current paper is not a survey of the method of \cite{BukhKlib}, then we
refer here only to a few such works \cite%
{BK,ImYam1,Isakov,Klib84,Klib92,KT,Ksurvey,KL,Yam}. The idea of \cite%
{BukhKlib} has also found its applications in the so-called convexification
globally convergent numerical method for CIPs, see, e.g. \cite{KL} for main
numerical results as of 2021. The main advantage of the convexification over
conventional numerical methods for CIPs is that the well known phenomenon of
local minima and ravines of conventional least squares cost functionals does
not occur in the convexification.

From the PDE standpoint, we analyze the second order MFGS consisting of
Bellman and Fokker-Planck equations supplied with the zero Neumann boundary
condition, terminal condition for the Bellman equation and initial and
terminal conditions for the Fokker-Planck equation.

The mean field games theory studies the mass behavior of identical rational
agents. This approach goes back to the notion of the Walrasian equilibrium
as well as to the models of statistical physics and presumes the study of
infinitely many players optimizing their own payoff depending on the
dynamics of the whole mass in the case when a single player does not affect
the dynamics of the whole mass. Among model systems examined within the mean
field games theory are various models of finance, economics, pedestrians
flocking, interactions of electric vehicles \cite{Cou}, etc.

The mean field games theory (MFG) was first introduced by Lasry and Lions 
\cite{Lasry_Lions_2006_I,Lasry_Lions_2006_II} and Huang, Caines and Malham%
\'{e} \cite{Huang_Caines_Malhame_2007,Huang_Malhame_Caines_2006}. Those
pioneering publications first reduced the game theoretical setting to a pair
of PDEs consisting of Bellman and Fokker-Planck equations. Recall that the
Bellman equation describes the value function for a representative player,
while the Fokker-Planck equation provides the dynamics of the distribution
of all players. The classical setting of the MFG implies that the terminal
condition for the Bellman equation and the initial condition for the
Fokker-Planck equation are given. The first condition is the terminal payoff
for the representative player, while the second condition gives the initial
distribution of players. We also note that the equilibrium strategy can be
designed based on a solution of the Bellman equation. This enables one to
solve the infinite player game based on a solution of the MFGS.

The first publications about MFG dealt with the case when the dynamics of
each agent is given by a simple controlled stochastic differential equation,
while the payoff is the sum of the energy term with the interaction term.
Nowadays, the existence of the solution of the MFGS is proved for general
dynamics that can include drift, Brownian motion and jumps \cite%
{Kolokoltsov_Yang_2013_existence}. We note that the existence result can be
obtained either through an accurate analysis of the MFGS or using the
so-called probabilistic approach that goes back to the infinite player game
and proves the existence of the Nash equilibrium for the infinite player
game \cite{Carmona_Delarue_I}, \cite{Carmona_Delarue_II}.

On the other hand, uniqueness theorems for the classical setting of the MFGS
are more rare. Indeed, in general, the Nash equilibrium is non-unique. This
property is inherited by the MFGS. Therefore, the uniqueness of the solution
to the MFGS is proved only under some additional constraints, like, e.g.
monotonicity assumptions \cite{Bardi_Fischer_2017}. Stability estimates for
the solutions of MFGS with respect to the possible noise in initial and
terminal data are unknown.

A different setting is the so-called planning problem, which assumes that
only conditions on the Fokker-Planck equation are given. From the
game-theoretical standpoint, the planning problem means that one is trying
to find a terminal payoff for a representative player such that the
corresponding solution of the MFGS would lead the whole mass of players from
the given initial distribution to the prescribed final one. The results of
papers \cite{Porrtte}, \cite{Graber_et_al_2019} provide the existence and
uniqueness of the solution of the planning problem for the case of local
coupling and monotone interaction term. However, there are no general
existence results. Moreover, it is shown for the toy model of a finite state
mean field game that the solution of the planning problem might not exist 
\cite{Averboukh_Volkov}.

Below $x\in \mathbb{R}^{n}$\ denotes the position $x$\ of an agent and $%
t\geq 0$ is time. Let $T>0$ be a number and $\Omega \subset \mathbb{R}^{n}$
be a bounded domain with a piecewise smooth boundary $\partial \Omega $.
Denote 
\begin{equation*}
Q_{T}=\Omega \times \left( 0,T\right) ,S_{T}=\partial \Omega \times \left(
0,T\right) .
\end{equation*}%
Let $v(x,t)$\ be the value function and $m(x,t)$\ be the distribution of
agents at the point $x$\ and at the moment of time $t.$ We study in this
paper the second order mean field games system

\begin{equation}
\left. 
\begin{array}{c}
v_{t}(x,t)+\beta \Delta v(x,t)+\varkappa ^{2}(x)(\nabla v(x,t))^{2}/2+ \\ 
+F\left( x,t,\dint\limits_{\Omega }K\left( x,y\right) m\left( y,t\right)
dy,m\left( x,t\right) \right) =0,\text{ }\left( x,t\right) \in Q_{T}, \\ 
m_{t}(x,t)-\beta \Delta m(x,t)+ \\ 
+\nabla \cdot (\varkappa ^{2}(x)m(x,t)\nabla v(x,t))=0,\text{ }\left(
x,t\right) \in Q_{T}.%
\end{array}%
\right.  \label{1.1}
\end{equation}%
We refer to Appendix (section 5) for a brief outline of a derivation of this
system.

The meaning of the function $\varkappa (x)$\ is explained in section 5. The
function $F$ is an interaction term that combines the nonlocal interaction
given by the integral operator as well as local interaction through the
fourth argument of $F$, also, see section 5. Thus, it follows from (\ref{1.1}%
) that we consider a general case when both local and non-local interactions
are included in the interaction function $F$.

We assume the zero Neumann boundary conditions 
\begin{equation}
\begin{split}
& \partial _{n}v(x,t)=0,\ \ \left( x,t\right) \in S_{T}, \\
& \partial _{n}m(x,t)=0,\ \ \left( x,t\right) \in S_{T},
\end{split}
\label{1.2}
\end{equation}%
where $n\left( x\right) $ is the outward looking normal vector at $\partial
\Omega .$ Conditions (\ref{1.2}) mean that the flux through the boundary of
both the value function and agents is zero. In addition, we assume that the
terminal condition is given for the function $v\left( x,t\right) $ and both
initial and terminal conditions are given for the function $m\left(
x,t\right) $:%
\begin{equation}
\left. 
\begin{array}{c}
v\left( x,T\right) =v_{T}\left( x\right) , \\ 
m\left( x,0\right) =m_{0}\left( x\right) , \\ 
m\left( x,T\right) =m_{T}\left( x\right) , \\ 
v_{T}\left( x\right) ,m_{0}\left( x\right) ,m_{T}\left( x\right) \in
H^{1}\left( \Omega \right) .%
\end{array}%
\right.  \label{1.3}
\end{equation}

Problem (\ref{1.1})-(\ref{1.3}) is an overdetermined mean field game. The
overdetermination comes from the knowledge of the function $m_{T}\left(
x\right) .$ In practical settings the functions in the right hand sides of
equalities (\ref{1.3}) might be given with a noise/error. Therefore, the
question can be raised on how the noise affects the solution of problem (\ref%
{1.1})-(\ref{1.3}). Thus, we prove below Lipschitz stability estimates with
respect to that noise for both functions $v$ and $m.$

It is well-known that if we would fix an initial distribution of players,
then we would arrive at the MFGS (\ref{1.1}) with the initial condition
imposed on $m\left( x,t\right) $ and the terminal condition imposed on $%
v\left( x,t\right) $. The setting of the overdetermined problem appears when
one observes a system of infinitely many rational players, and one thinks
that this system exhibits an equilibrium behavior. Our first goal here is to
demonstrate that this equilibrium, if it exists, is stable with respect to
the input data: initial and terminal data for the function $m\left(
x,t\right) $ and terminal data for the function $v\left( x,t\right) .$ Our
second goal is to prove that, if these data are given, then this
equilibrium, if it exists, is unique. Note that such a setting, implicitly
appeared in \cite{Trusov} where the real data from the Chinese stock market
crash in 2015 where analyzed using the mean field games. In addition, the
retrospective problem is reasonable for the analysis of the pedestrian
flocking models.

Certainly, if the MFGS has only one solution, then this solution can be
found from the usual initial and terminal conditions. However, as it was
mentioned above, generally the MFGS has multiple solutions. Thus, to figure
out which solution was actually realized in the process, we need an
additional information. It is reasonable to assume that the final
distribution $m\left( x,T\right) =m_{T}\left( x\right) $ can be measured.
This assumption leads to the overdetermined problem (\ref{1.1})-(\ref{1.3})
for the MFGS. As it was mentioned above, we study the questions of stability
and uniqueness questions of the solution of this problem. Since a noise is
always present in real measurements, then it makes sense to estimate the
error of the solution of that overdetermined problem for MFGS with respect
to the level of noise in the data. We call this \textquotedblleft stability
estimate".

As to the possible further developments of the ideas of this paper, the tool
of Carleman estimates for MFGS (\ref{1.1}), which is used here, seems to be
quite helpful for Coefficient Inverse Problems (CIPs) for this system.
Indeed, this paper is a revised version of the preprint \cite{MFG1}. After 
\cite{MFG1} was posted on www.arxiv.org on February 21, 2023, the first
author with coauthors posted/published two works \cite{MFG5,MFG6}, in which
Carleman estimates are applied to prove stability and uniqueness results for
CIPs for MFGS (\ref{1.1}). The framework of \cite{BukhKlib} is used in these
references. In \cite{MFG5} CIPs with the final observation are considered.
In \cite{MFG6} the data for the CIP are given at $\left\{ t=t_{0}\right\} ,$
where $t_{0}\in \left( 0,T\right) $ is a fixed number, and they are
complemented by the lateral Cauchy data.

We now refer to previous publications about CIPs for the MFGS. To the best
of our knowledge, the first theoretical results about CIPs for MFGS (\ref%
{1.1}) were obtained in \cite{Liu1,Liu2}. In these works, uniqueness
theorems are proven about the recovery of either the function $F$ in the
first equation (\ref{1.1}) \cite{Liu1} or of both functions $F$ and $%
G=G\left( x,m\left( x,T\right) \right) =u\left( x,T\right) $ \cite{Liu2}
from the data, which depend on the function $m_{0}\left( x\right) ,$ and
this function varies over certain sets of functions$.$In \cite{Liu1} $%
F=F\left( x,m\left( x,t\right) \right) $ and in \cite{Liu2} $F=F\left(
x,t,m\left( x,t\right) \right) .$ Since the function $m_{0}\left( x\right) $
varies, then the data in \cite{Liu1,Liu2} represent multiple measurements.
On the other hand, the data in our paper as well as in \cite{MFG5,MFG6} are
results of a single measurement event. Numerical studies of CIPs for the
MFGS were conducted in \cite{Chow,Ding}.

Everywhere below $\beta =const.>0$ and all functions are real valued ones.
In section 2 we prove two integral identities. In section 3 we prove two new
Carleman estimates. In section 4 we prove the Lipschitz stability estimate
and uniqueness for problem (\ref{1.1})-(\ref{1.3}). In section 5 we provide
a brief derivation of system (\ref{1.1}).

\section{Two Integral Identities}

\label{sec:2}

In this section, we prove two integral identities for $\left\Vert \Delta
u\right\Vert _{L_{2}\left( \Omega \right) }^{2}$ for an arbitrary function $%
u\in H^{2}\left( \Omega \right) $ satisfying either zero Neumann or zero
Dirichlet boundary condition at $\partial \Omega $. A similar estimate for a
general elliptic operator of the second order with variable coefficients is
proven in \cite[Chapter 2, \S 6]{Lad} for the case when the function $u$
satisfies zero Dirichlet boundary condition at $\partial \Omega $.
Furthermore, Remark 6.2 in \cite[Chapter 2, \S 6]{Lad} states that this
estimate is also valid for the case of the zero Robin boundary condition,
which, in particular, includes the zero Neumann boundary condition. This
remark is not proven in \cite{Lad}.

\textbf{Remarks 2.1:}

\begin{enumerate}
\item \emph{We assume in Lemmata 2.1, 2.2 that we work with the case when
the domain }$\Omega $\emph{\ is a rectangular prism. Although we conjecture
that these lemmata can be extended to the case of a more general domain }$%
\Omega ,$\emph{\ we do not yet know how to do this. Besides, we use Lemma
2.1 only to strengthen somewhat our Lipschitz stability estimate of Theorem
4.1 since a certain Lipschitz stability estimate is valid without this lemma
anyway, see the formulation of Theorem 4.1. Thus, this lemma is of a
secondary importance for the specific goal of the current paper. }

\item \emph{As to Lemma 2.2, we do not use it below and bring it in here
only for the sake of completeness.}

\item \emph{We believe that Lemmata 2.1, 2.2 are interesting in their own
rights since these are integral identities rather than the classical
estimate of \cite[Chapter 2, \S 6]{Lad}.}
\end{enumerate}

Let $A_{1},...,A_{n}>0$ be some numbers. We consider in this section the
case when $\Omega \subset \mathbb{R}^{n}$ is a rectangular prism,%
\begin{equation}
\Omega =\left\{ x=\left( x_{1},...,x_{n}\right)
:-A_{i}<x_{i}<A_{i},i=1,...,n\right\} .  \label{4.1}
\end{equation}%
For $i=1,...,n$ denote: 
\begin{equation}
\Gamma _{i}^{+}=\left\{ x_{i}=A_{i},-A_{j}<x_{j}<A_{j},1\leq j\leq n,j\neq
i\right\} ,  \label{4.100}
\end{equation}%
\begin{equation}
\Gamma _{i}^{-}=\left\{ x_{i}=-A_{i},-A_{j}<x_{j}<A_{j},1\leq j\leq n,j\neq
i\right\} .  \label{4.101}
\end{equation}%
By (\ref{4.1})-(\ref{4.101})%
\begin{equation}
\partial \Omega =\left( \cup _{i=1}^{n}\Gamma _{i}^{+}\right) \cup \left(
\cup _{i=1}^{n}\Gamma _{i}^{-}\right) .  \label{4.102}
\end{equation}

\textbf{Lemma 2.1.} \emph{The following integral identity holds:}%
\begin{equation}
\dint\limits_{\Omega }\left( \Delta u\right)
^{2}dx=\dsum\limits_{i,j=1}^{n}\dint\limits_{\Omega }u_{x_{i}x_{j}}^{2}dx,
\label{4.2}
\end{equation}%
\begin{equation}
\forall u\in \left\{ u\in H^{2}\left( \Omega \right) :\partial _{n}u\mid
_{\Omega }=0\right\} .  \label{4.3}
\end{equation}

\textbf{Proof.} Temporary assume that 
\begin{equation}
u\in \left\{ u:u\in C^{3}\left( \overline{\Omega }\right) ,\partial
_{n}u\mid _{\Omega }=0\right\} .  \label{4.4}
\end{equation}

We have: 
\begin{equation*}
\left( \Delta u\right)
^{2}=\dsum\limits_{i=1}^{n}u_{x_{i}x_{i}}^{2}+\dsum\limits_{i,j=1,i\neq
j}^{n}u_{x_{i}x_{i}}u_{x_{j}x_{j}}=
\end{equation*}%
\begin{equation*}
=\dsum\limits_{i=1}^{n}u_{x_{i}x_{i}}^{2}+\dsum\limits_{i,j=1,i\neq
j}^{n}\left( u_{x_{i}x_{i}}u_{x_{j}}\right)
_{x_{j}}-\dsum\limits_{i,j=1,i\neq j}^{n}u_{x_{j}x_{i}x_{i}}u_{x_{j}}=
\end{equation*}%
\begin{equation*}
=\dsum\limits_{i=1}^{n}u_{x_{i}x_{i}}^{2}+\dsum\limits_{i,j=1,i\neq
j}^{n}\left( u_{x_{i}x_{i}}u_{x_{j}}\right)
_{x_{j}}+\dsum\limits_{i,j=1,i\neq j}^{n}\left(
-u_{x_{i}x_{j}}u_{x_{j}}\right) _{x_{i}}+\dsum\limits_{i,j=1,i\neq
j}^{n}u_{x_{i}x_{j}}^{2}=
\end{equation*}%
\begin{equation*}
=\dsum\limits_{i,j=1}^{n}u_{x_{i}x_{j}}^{2}+\dsum\limits_{i,j=1,i\neq
j}^{n}\left( u_{x_{i}x_{i}}u_{x_{j}}\right)
_{x_{j}}+\dsum\limits_{i,j=1,i\neq j}^{n}\left(
-u_{x_{i}x_{j}}u_{x_{j}}\right) _{x_{i}}.
\end{equation*}%
Thus,%
\begin{equation}
\left( \Delta u\right)
^{2}=\dsum\limits_{i,j=1}^{n}u_{x_{i}x_{j}}^{2}+\dsum\limits_{i,j=1,i\neq
j}^{n}\left( u_{x_{i}x_{i}}u_{x_{j}}\right)
_{x_{j}}+\dsum\limits_{i,j=1,i\neq j}^{n}\left(
-u_{x_{i}x_{j}}u_{x_{j}}\right) _{x_{i}}.  \label{4.5}
\end{equation}%
Integrating identity (\ref{4.5}) over $\Omega $, using Gauss formula and (%
\ref{4.1})-(\ref{4.102}), we obtain%
\begin{equation*}
\dint\limits_{\Omega }\left( \Delta u\right)
^{2}dx=\dsum\limits_{i,j=1}^{n}\dint\limits_{\Omega }u_{x_{i}x_{j}}^{2}dx+
\end{equation*}%
\begin{equation}
+\dsum\limits_{i,j=1,i\neq j}^{n}\left( \dint\limits_{\Gamma _{j}^{+}}\left(
u_{x_{i}x_{i}}u_{x_{j}}\right) \left( x\right) dS-\dint\limits_{\Gamma
_{j}^{-}}\left( u_{x_{i}x_{i}}u_{x_{j}}\right) \left( x\right) dS\right) +
\label{4.6}
\end{equation}%
\begin{equation*}
\dsum\limits_{i,j=1,i\neq j}^{n}\left( -\dint\limits_{\Gamma _{i}^{+}}\left(
u_{x_{i}x_{j}}u_{x_{j}}\right) \left( x\right) dS+\dint\limits_{\Gamma
_{i}^{-}}\left( u_{x_{i}x_{i}}u_{x_{j}}\right) \left( x\right) dS\right) .
\end{equation*}%
We now prove that all terms in the second and third lines of (\ref{4.6})
equal to zero. Consider for example the term with $j=1$ in the second line
of (\ref{4.6}). This term is:%
\begin{equation*}
\dsum\limits_{i=2}^{n}\dint\limits_{\Gamma _{1}^{+}}\left(
u_{x_{i}x_{i}}u_{x_{1}}\right) \left( A_{1},x_{2},...,x_{n}\right)
dx_{2}...dx_{n}-
\end{equation*}%
\begin{equation}
-\dsum\limits_{i=2}^{n}\dint\limits_{\Gamma _{1}^{-}}\left(
u_{x_{i}x_{i}}u_{x_{1}}\right) \left( -A_{1},x_{2},...,x_{n}\right)
dx_{2}...dx_{n}.  \label{4.7}
\end{equation}%
Since by (\ref{4.1})-(\ref{4.102}) and (\ref{4.4}) 
\begin{equation}
u_{x_{1}}\left( A_{1},x_{2},...,x_{n}\right) =u_{x_{1}}\left(
-A_{1},x_{2},...,x_{n}\right) =0,  \label{4.8}
\end{equation}
then both lines of (\ref{4.7}) equal to zero. Similarly, the same is true
for all other terms in the second line of (\ref{4.6}). Hence, the second
line of (\ref{4.6}) equals zero, 
\begin{equation}
\dsum\limits_{i,j=1,i\neq j}^{n}\left( \dint\limits_{\Gamma _{j}^{+}}\left(
u_{x_{i}x_{i}}u_{x_{j}}\right) \left( x\right) dS-\dint\limits_{\Gamma
_{j}^{-}}\left( u_{x_{i}x_{i}}u_{x_{j}}\right) \left( x\right) dS\right) =0.
\label{4.9}
\end{equation}

We are now concerned with the third line of (\ref{4.6}). Consider, for
example the term with $i=1$. This term is:%
\begin{equation*}
-\dsum\limits_{j=2}^{n}\dint\limits_{\Gamma _{1}^{+}}\left(
u_{x_{1}x_{j}}u_{x_{j}}\right) \left( A_{1},x_{2},...,x_{n}\right)
dx_{2}...dx_{n}+
\end{equation*}%
\begin{equation}
+\dsum\limits_{j=2}^{n}\dint\limits_{\Gamma _{1}^{+}}\left(
u_{x_{1}x_{j}}u_{x_{j}}\right) \left( -A_{1},x_{2},...,x_{n}\right)
dx_{2}...dx_{n}.  \label{4.10}
\end{equation}%
By (\ref{4.8})%
\begin{equation}
u_{x_{1}x_{j}}\left( A_{1},x_{2},...,x_{n}\right) =u_{x_{1}x_{j}}\left(
-A_{1},x_{2},...,x_{n}\right) =0,j=2,...,n.  \label{4.200}
\end{equation}%
Hence, both lines of (\ref{4.10}) equal to zero. Similarly we obtain the
same conclusion for all other terms of the third line of (\ref{4.6}). Hence,
the third line of (\ref{4.6}) equals to zero,%
\begin{equation}
\dsum\limits_{i,j=1,i\neq j}^{n}\left( -\dint\limits_{\Gamma _{i}^{+}}\left(
u_{x_{i}x_{j}}u_{x_{j}}\right) \left( x\right) dS+\dint\limits_{\Gamma
_{i}^{-}}\left( u_{x_{i}x_{i}}u_{x_{j}}\right) \left( x\right) dS\right) =0.
\label{4.11}
\end{equation}%
Equalities (\ref{4.6}), (\ref{4.9}) and (\ref{4.11}) imply the integral
identity (\ref{4.2}) for all functions $u$ satisfying (\ref{4.4}). Using
density arguments, we obtain that (\ref{4.2}) is valid for all functions $u$
satisfying (\ref{4.3}). $\square $

\textbf{Lemma 2.2. }\emph{The integral identity (\ref{4.2}) remains valid if
the zero Neumann boundary condition (\ref{4.3}) is replaced with the zero
Dirichlet boundary condition, }%
\begin{equation}
u\in \left\{ u\in H^{2}\left( \Omega \right) :u\mid _{\partial \Omega
}=0\right\} .  \label{4.12}
\end{equation}

\textbf{Proof.} Consider, for example again the case when $j=1$ in the
second line of (\ref{4.6}). This term is presented in (\ref{4.7}). Since in (%
\ref{4.7}) $i\neq 1,$ then by (\ref{4.12})%
\begin{equation*}
u_{x_{i}x_{i}}\left( A_{1},x_{2},...,x_{n}\right) =u_{x_{i}x_{i}}\left(
-A_{1},x_{2},...,x_{n}\right) =0.
\end{equation*}%
Hence, both lines of (\ref{4.7}) equal to zero. Similarly, the same is true
for all other terms in the second line of (\ref{4.6}). Hence, the second
line of (\ref{4.6}) equals zero, i.e. (\ref{4.9}) holds.

As to the third line of (\ref{4.6}), consider again, for example the case $%
i=1$ in it. The corresponding term is as in (\ref{4.10}). It follows from (%
\ref{4.12}) that (\ref{4.200}) holds. Hence, (\ref{4.11}) holds as well. $%
\square $

\section{Two Carleman Estimates}

\label{sec:3}

Any Carleman estimate is dependent only on the principal part of a PDE
operator and is independent on its low order terms, see, e.g. \cite[Lemma
2.1.1]{KL}. Thus, we derive in this section two new Carleman estimates only
for the operators $\partial _{t}+\beta \Delta $ and $\partial _{t}-\beta
\Delta $ of the first and second equations (\ref{1.1}) respectively.

Introduce the subspace $H_{0}^{2}\left( Q_{T}\right) $ of the space $%
H^{2}\left( Q_{T}\right) $ as:%
\begin{equation}
H_{0}^{2}\left( Q_{T}\right) =\left\{ u\in H^{2}\left( Q_{T}\right)
:\partial _{n}u\mid _{S_{T}}=0\right\} .  \label{2.00}
\end{equation}%
Note that by trace theorem 
\begin{equation}
u\left( x,0\right) ,u\left( x,T\right) \in H^{1}\left( \Omega \right) ,\text{
}\forall u\in H_{0}^{2}\left( Q_{T}\right) .  \label{2.000}
\end{equation}%
Let $a>1$ be a number. Everywhere below $C=C\left( T,a\right) >0$ denotes
different positive numbers depending only on listed parameters.\emph{\ }Let $%
\lambda >0$ and $\nu >2$ be two parameters, which we will choose later.
Consider the Carleman Weight Function $\varphi _{\lambda ,\nu }\left(
t\right) ,$%
\begin{equation}
\varphi _{\lambda ,\nu }\left( t\right) =e^{2\lambda \left( t+a\right) ^{\nu
}},t\in \left( 0,T\right) .  \label{2.0}
\end{equation}%
This function is used as the weight function in our two Carleman estimates.

\subsection{The first Carleman estimate}

\label{sec:3.1}

\textbf{Theorem 3.1.} \emph{There exists a number }$C=C\left( T,a\right) >0$%
\emph{\ depending only on listed parameters such that the following Carleman
estimate is valid:}%
\begin{equation}
\left. 
\begin{array}{c}
\dint\limits_{Q_{T}}\left( u_{t}+\beta \Delta u\right) ^{2}\varphi _{\lambda
,\nu }^{2}dxdt\geq \dint\limits_{Q_{T}}\left( u_{t}^{2}/4+\beta ^{2}\left(
\Delta u\right) ^{2}\right) \varphi _{\lambda ,\nu }^{2}dxdt+ \\ 
+C\lambda \nu \beta \dint\limits_{Q_{T}}\left( \nabla u\right) ^{2}\varphi
_{\lambda ,\nu }^{2}dxdt+C\lambda ^{2}\nu
^{2}\dint\limits_{Q_{T}}u^{2}\varphi _{\lambda ,\nu }^{2}dxdt- \\ 
-e^{2\lambda \left( T+a\right) ^{\nu }}\dint\limits_{\Omega }\left[ \beta
\left( \nabla _{x}u\right) ^{2}+\lambda \nu \left( T+a\right) ^{\nu -1}u^{2}%
\right] \left( x,T\right) dt, \\ 
\forall \lambda >0,\forall \nu >2,\forall u\in H_{0}^{2}\left( Q_{T}\right) .%
\end{array}%
\right.  \label{2.1}
\end{equation}%
\emph{In particular, if the domain }$\Omega $\emph{\ is as the one in (\ref%
{4.1}), then (\ref{2.1}) becomes:}%
\begin{equation}
\left. 
\begin{array}{c}
\dint\limits_{Q_{T}}\left( u_{t}+\beta \Delta u\right) ^{2}\varphi _{\lambda
,\nu }^{2}dxdt\geq \dint\limits_{Q_{T}}\left( u_{t}^{2}/4+\beta
^{2}\dsum\limits_{i,j=1}^{n}u_{x_{i}x_{j}}^{2}\right) \varphi _{\lambda ,\nu
}^{2}dxdt+ \\ 
+C\lambda \nu \dint\limits_{Q_{T}}\left( \nabla _{x}u\right) ^{2}\varphi
_{\lambda ,\nu }^{2}dxdt+C\lambda ^{2}\nu
^{2}\dint\limits_{Q_{T}}u^{2}\varphi _{\lambda ,\nu }^{2}dxdt- \\ 
-e^{2\lambda \left( T+a\right) ^{\nu }}\dint\limits_{\Omega }\left[ \beta
\left( \nabla _{x}u\right) ^{2}+\lambda \nu \left( T+a\right) ^{\nu -1}u^{2}%
\right] \left( x,T\right) dt, \\ 
\forall \lambda >0,\forall \nu >2,\forall u\in H_{0}^{2}\left( Q_{T}\right) .%
\end{array}%
\right.  \label{2.100}
\end{equation}

\textbf{Proof.} In this proof $u\in H^{2}\left( Q_{T}\right) $ is an
arbitrary function$.$ Thus, (\ref{2.000}) implies that integrals in the
third line of (\ref{2.1}) make sense.

Introduce a new function $w\left( x,t\right) ,$ 
\begin{equation}
w\left( x,t\right) =u\left( x,t\right) e^{\lambda \left( t+a\right) ^{\nu }}.
\label{2.2}
\end{equation}%
Then%
\begin{equation}
\left. 
\begin{array}{c}
u=we^{-\lambda \left( t+a\right) ^{\nu }}, \\ 
u_{t}=\left( w_{t}-\lambda \nu \left( t+a\right) ^{\nu -1}w\right)
e^{-\lambda \left( t+a\right) ^{\nu }}, \\ 
\Delta u=\Delta we^{-\lambda \left( t+a\right) ^{\nu }}.%
\end{array}%
\right.  \label{2.20}
\end{equation}%
Hence,%
\begin{equation}
\left. 
\begin{array}{c}
\left( u_{t}+\beta \Delta u\right) ^{2}\varphi _{\lambda ,\nu }= \\ 
=\lambda ^{2}\nu ^{2}\left( t+a\right) ^{2\nu -2}w^{2}- \\ 
-2\lambda \nu \left( t+a\right) ^{\nu -1}w\left( w_{t}+\beta \Delta w\right)
+ \\ 
+\left( w_{t}+\beta \Delta w\right) ^{2}.%
\end{array}%
\right.  \label{2.3}
\end{equation}%
We now estimate from the below terms in the third and fourth lines of (\ref%
{2.3}).

\textbf{Step 1.} Estimate sum of the second and third lines of (\ref{2.3})
from the below. We have for the third line: 
\begin{equation*}
-2\lambda \nu \left( t+a\right) ^{\nu -1}w\left( w_{t}+\beta \Delta w\right)
=-2\lambda \nu \left( t+a\right) ^{\nu -1}ww_{t}-2\lambda \nu \left(
t+a\right) ^{\nu -1}\beta w\Delta w=
\end{equation*}%
\begin{equation*}
=\left( -\lambda \nu \left( t+a\right) ^{\nu -1}w^{2}\right) _{t}+\lambda
\nu \left( \nu -1\right) \left( t+a\right) ^{\nu -2}w^{2}+
\end{equation*}%
\begin{equation*}
+\dsum\limits_{i=1}^{n}\left( -2\lambda \nu \left( t+a\right) ^{\nu -1}\beta
ww_{x_{i}}\right) _{x_{i}}+2\lambda \nu \left( t+a\right) ^{\nu -1}\beta
\left( \nabla w\right) ^{2}.
\end{equation*}%
Combining this with the second line of (\ref{2.3}), we obtain the following
estimate for the sum of second and third lines of of (\ref{2.3}): 
\begin{equation*}
\lambda ^{2}\nu ^{2}\left( t+a\right) ^{2\nu -2}w^{2}-2\lambda \nu \left(
t+a\right) ^{\nu -1}w\left( w_{t}+\beta \Delta w\right) \geq
\end{equation*}%
\begin{equation*}
\geq \lambda ^{2}\nu ^{2}\left( t+a\right) ^{2\nu -2}w^{2}+2\lambda \nu
\left( t+a\right) ^{\nu -1}\beta \left( \nabla w\right) ^{2}+
\end{equation*}%
\begin{equation}
+\left( -\lambda \nu \left( t+a\right) ^{\nu -1}w^{2}\right)
_{t}+\dsum\limits_{i=1}^{n}\left( -2\lambda \nu \left( t+a\right) ^{\nu
-1}\beta ww_{x_{i}}\right) _{x_{i}}.  \label{2.4}
\end{equation}

\textbf{Step 2.} Consider the fourth line of (\ref{2.3}),%
\begin{equation*}
\left( w_{t}+\beta \Delta w\right) ^{2}=w_{t}^{2}+2\beta w_{t}\Delta w+\beta
^{2}\left( \Delta w\right) ^{2}=
\end{equation*}%
\begin{equation*}
=w_{t}^{2}+2\beta \dsum\limits_{i=1}^{n}w_{t}w_{x_{i}x_{i}}+\beta ^{2}\left(
\Delta w\right) ^{2}=
\end{equation*}%
\begin{equation*}
=w_{t}^{2}+\dsum\limits_{i=1}^{n}\left( 2\beta w_{t}w_{x_{i}}\right)
_{x_{i}}-2\beta \dsum\limits_{i=1}^{n}w_{tx_{i}}w_{x_{i}}+\beta ^{2}\left(
\Delta w\right) ^{2}=
\end{equation*}%
\begin{equation*}
=w_{t}^{2}+\beta ^{2}\left( \Delta w\right)
^{2}+\dsum\limits_{i=1}^{n}\left( 2\beta w_{t}w_{x_{i}}\right)
_{x_{i}}+\left( -\beta \left( \nabla w\right) ^{2}\right) _{t}.
\end{equation*}

Thus, 
\begin{equation}
\left( w_{t}+\beta \Delta w\right) ^{2}=w_{t}^{2}+\beta ^{2}\left( \Delta
w\right) ^{2}+\dsum\limits_{i=1}^{n}\left( 2\beta w_{t}w_{x_{i}}\right)
_{x_{i}}+\left( -\beta \left( \nabla w\right) ^{2}\right) _{t}.  \label{2.5}
\end{equation}

\textbf{Step 3.} Combine (\ref{2.3})-(\ref{2.5}). We obtain%
\begin{equation*}
\left( u_{t}+\beta \Delta u\right) ^{2}\varphi _{\lambda ,\nu }^{2}\left(
t\right) \geq \lambda ^{2}\nu ^{2}\left( t+a\right) ^{2\nu -2}w^{2}+2\lambda
\nu \left( t+a\right) ^{\nu -1}\beta \left( \nabla w\right) ^{2}+
\end{equation*}%
\begin{equation}
+w_{t}^{2}+\beta ^{2}\left( \Delta w\right) ^{2}+  \label{2.6}
\end{equation}%
\begin{equation*}
+\left( -\lambda \nu \left( t+a\right) ^{\nu -1}w^{2}-\beta \left( \nabla
w\right) ^{2}\right) _{t}+
\end{equation*}%
\begin{equation*}
+\dsum\limits_{i=1}^{n}\left( -2\lambda \nu \left( t+a\right) ^{\nu -1}\beta
ww_{x_{i}}+2\beta w_{t}w_{x_{i}}\right) _{x_{i}}.
\end{equation*}%
Next, using (\ref{2.2}) and Cauchy-Schwarz inequality, we obtain%
\begin{equation*}
w_{t}^{2}\geq \frac{1}{2}w_{t}^{2}=\frac{1}{2}\left( u_{t}^{2}+2\lambda \nu
\left( t+a\right) ^{\nu -1}u_{t}u+\lambda ^{2}\nu ^{2}\left( t+a\right)
^{2\nu -2}u^{2}\right) \varphi _{\lambda ,\nu }\left( t\right) \geq
\end{equation*}%
\begin{equation*}
\geq \frac{1}{2}\left( \frac{1}{2}u_{t}^{2}-\lambda ^{2}\nu ^{2}\left(
t+a\right) ^{2\nu -2}u^{2}\right) \varphi _{\lambda ,\nu }\left( t\right) .
\end{equation*}%
Hence, combining this with (\ref{2.6}) and using 
\begin{equation}
\left( t+a\right) ^{\nu -1}>1,t\in \left( 0,T\right) ,  \label{2.70}
\end{equation}%
we obtain%
\begin{equation*}
\left( u_{t}+\beta \Delta u\right) ^{2}\varphi _{\lambda ,\nu }\geq C\left(
\lambda ^{2}\nu ^{2}u^{2}+\lambda \nu \beta \left( \nabla u\right)
^{2}\right) \varphi _{\lambda ,\nu }+
\end{equation*}%
\begin{equation}
+\left( \frac{1}{4}u_{t}^{2}+\beta ^{2}\left( \Delta u\right) ^{2}\right)
\varphi _{\lambda ,\nu }+  \label{2.7}
\end{equation}%
\begin{equation*}
+\left[ \left( -\lambda \nu \left( t+a\right) ^{\nu -1}u^{2}-\beta \left(
\nabla u\right) ^{2}\right) \varphi _{\lambda ,\nu }\right] _{t}+
\end{equation*}%
\begin{equation*}
+\dsum\limits_{i=1}^{n}\left( -2\lambda \nu \left( t+a\right) ^{\nu -1}\beta
ww_{x_{i}}+2\beta w_{t}w_{x_{i}}\right) _{x_{i}}.
\end{equation*}%
Integrate estimate (\ref{2.7}) over the domain $Q_{T}.$ Since $\partial
_{n}u\mid _{S_{T}}=0,$ then, by Gauss formula, the integral over $Q_{T}$ of
the last line of (\ref{2.7}) equals zero. Thus, we obtain the first target
estimate (\ref{2.1}). The second target estimate (\ref{2.100}) follows
immediately from a combination of (\ref{2.1}) with Lemma 2.1. $\square $

\subsection{The second Carleman estimate}

\label{sec:3.2}

\textbf{Theorem 3.2.} \emph{There exist a sufficiently large number }$\nu
_{0}=\nu _{0}\left( \beta ,T,a\right) >2$\emph{\ and a number }$C=C\left(
T,a\right) >0$\emph{\ depending only on listed parameters such that the
following Carleman estimate holds:} 
\begin{equation}
\left. 
\begin{array}{c}
\dint\limits_{Q_{T}}\left( u_{t}-\beta \Delta u\right) ^{2}\varphi _{\lambda
,\nu }dxdt\geq \\ 
\geq C\beta \sqrt{\nu }\dint\limits_{Q_{T}}\left( \nabla u\right)
^{2}\varphi _{\lambda ,\nu }dxdt+C\lambda \nu
^{2}\dint\limits_{Q_{T}}u^{2}\varphi _{\lambda ,\nu }dxdt- \\ 
-C\lambda \nu \left( T+a\right) ^{\nu -1}e^{2\lambda \left( T+a\right) ^{\nu
}}\dint\limits_{\Omega }u^{2}\left( x,T\right) dx- \\ 
-Ce^{2\lambda a^{\nu }}\dint\limits_{\Omega }\left[ \left( \nabla u\right)
^{2}+\sqrt{\nu }u^{2}\right] \left( x,0\right) dx, \\ 
\forall \lambda >0,\forall \nu \geq \nu _{0},\forall u\in H_{0}^{2}\left(
Q_{T}\right) .%
\end{array}%
\right.  \label{2.8}
\end{equation}

\textbf{Proof}. In this proof $u\in H_{0}^{2}\left( Q_{T}\right) $ is an
arbitrary function. Thus, again (\ref{2.000}) implies that integrals in the
third line of (\ref{2.8}) make sense.

Consider again the change of variables (\ref{2.2}) as well as follow up
formulas (\ref{2.20}). Then%
\begin{equation}
\left. 
\begin{array}{c}
\left( u_{t}-\beta \Delta u\right) ^{2}\varphi _{\lambda ,\nu }=\left[
w_{t}-\left( \lambda \nu \left( t+a\right) ^{\nu -1}w+\beta \Delta w\right) %
\right] ^{2}\geq \\ 
\geq -2w_{t}\left( \lambda \nu \left( t+a\right) ^{\nu -1}w+\beta \Delta
w\right) =\left( -\lambda \nu \left( t+a\right) ^{\nu -1}w^{2}\right) _{t}+
\\ 
+\lambda \nu \left( \nu -1\right) \left( t+a\right) ^{\nu
-2}w^{2}+\dsum\limits_{i=1}^{n}\left( -2\beta w_{t}w_{x_{i}}\right)
_{x_{i}}+2\beta \nabla w_{t}\nabla w= \\ 
=\lambda \nu \left( \nu -1\right) \left( t+a\right) ^{\nu -2}w^{2}+\left(
-\lambda \nu \left( t+a\right) ^{\nu -1}w^{2}+\beta \left( \nabla w\right)
^{2}\right) _{t}+ \\ 
+\dsum\limits_{i=1}^{n}\left( -2\beta w_{t}w_{x_{i}}\right) _{x_{i}}.%
\end{array}%
\right.  \label{2.80}
\end{equation}%
Going back to the function $u$ via the first line of (\ref{2.20}),
integrating then (\ref{2.80}) over $Q_{T}$, using Gauss formula, the fact
that $\partial _{n}u\mid _{S_{T}}=0$ and also using $\left( t+a\right)
^{-1}>\left( T+a\right) ^{-1}$ for $t\in \left( 0,T\right) ,$ we obtain from
(\ref{2.80}) that there exists a sufficiently large number $\nu _{0}=\nu
_{0}\left( \beta ,T,a\right) >2$ such that%
\begin{equation}
\left. 
\begin{array}{c}
\dint\limits_{Q_{T}}\left( u_{t}-\beta \Delta u\right) ^{2}\varphi _{\lambda
,\nu }^{2}dxdt\geq C\lambda \nu ^{2}\dint\limits_{Q_{T}}u^{2}\left(
t+a\right) ^{\nu -1}\varphi _{\lambda ,\nu }dxdt- \\ 
-\lambda \nu \left( T+a\right) ^{\nu -1}e^{2\lambda \left( T+a\right) ^{\nu
}}\dint\limits_{\Omega }u^{2}\left( x,T\right) dx-\beta e^{2\lambda a^{\nu
}}\dint\limits_{\Omega }\left( \nabla u\right) ^{2}\left( x,0\right) dx, \\ 
\forall \lambda >0,\forall \nu \geq \nu _{0}>2,\forall u\in H_{0}^{2}\left(
Q_{T}\right) .%
\end{array}%
\right.  \label{2.9}
\end{equation}

To obtain the non-negative term with $\left( \nabla u\right) ^{2}\varphi
_{\lambda ,\nu }$ in the right hand side of (\ref{2.8}), consider the
expression $2\left( u_{t}-\beta \Delta u\right) u\varphi _{\lambda ,\nu }.$
Using (\ref{2.0}), we obtain 
\begin{equation*}
2\left( u_{t}-\beta \Delta u\right) u\varphi _{\lambda ,\nu }=\left(
u^{2}\varphi _{\lambda ,\nu }\right) _{t}-2\lambda \nu \left( t+a\right)
^{\nu -1}u^{2}\varphi _{\lambda ,\nu }+
\end{equation*}%
\begin{equation*}
+\dsum\limits_{i=1}^{n}\left( -2\beta uu_{x_{i}}\varphi _{\lambda ,\nu
}\right) _{x_{i}}+2\beta \left( \nabla u\right) ^{2}\varphi _{\lambda ,\nu }.
\end{equation*}%
Integrating this over $Q_{T},$ we obtain%
\begin{equation}
\left. 
\begin{array}{c}
2\dint\limits_{Q_{T}}\left( u_{t}-\beta \Delta u\right) u\varphi _{\lambda
,\nu }dxdt\geq 2\beta \dint\limits_{Q_{T}}\left( \nabla u\right) ^{2}\varphi
_{\lambda ,\nu }dxdt- \\ 
-C\lambda \nu \dint\limits_{Q_{T}}u^{2}\left( t+a\right) ^{\nu -1}\varphi
_{\lambda ,\nu }dxdt-e^{2\lambda a^{\nu }}\dint\limits_{\Omega }u^{2}\left(
x,0\right) dx.%
\end{array}%
\right.  \label{2.10}
\end{equation}%
Assuming that $\nu \geq \nu _{0}>2,$ multiply (\ref{2.10}) by $\sqrt{\nu }$
and sum up with (\ref{2.9}). Since $\nu \geq \nu _{0}$ and the number $\nu
_{0}>2$ is sufficiently large, then $\lambda \nu ^{2}>>\lambda \nu ^{3/2}.$
Hence, using (\ref{2.70}) again, we obtain%
\begin{equation*}
2\sqrt{\nu }\dint\limits_{Q_{T}}\left( u_{t}-\beta \Delta u\right) u\varphi
_{\lambda ,\nu }dxdt+\dint\limits_{Q_{T}}\left( u_{t}-\beta \Delta u\right)
^{2}\varphi _{\lambda ,\nu }dxdt\geq
\end{equation*}%
\begin{equation}
\geq 2\beta \sqrt{\nu }\dint\limits_{Q_{T}}\left( \nabla u\right)
^{2}\varphi _{\lambda ,\nu }dxdt+C\lambda \nu
^{2}\dint\limits_{Q_{T}}u^{2}\varphi _{\lambda ,\nu }dxdt-  \label{2.11}
\end{equation}%
\begin{equation*}
-\lambda \nu \left( T+a\right) ^{\nu -1}e^{2\lambda \left( T+a\right) ^{\nu
}}\dint\limits_{\Omega }u^{2}\left( x,T\right) dx-e^{2\lambda a^{\nu
}}\dint\limits_{\Omega }\left[ \beta \left( \nabla u\right) ^{2}+\sqrt{\nu }%
u^{2}\right] \left( x,0\right) dx.
\end{equation*}%
Next, 
\begin{equation}
2\sqrt{\nu }\dint\limits_{Q_{T}}\left( u_{t}-\beta \Delta u\right) u\varphi
_{\lambda ,\nu }dxdt+\dint\limits_{Q_{T}}\left( u_{t}-\beta \Delta u\right)
^{2}\varphi _{\lambda ,\nu }dxdt\leq  \label{2.12}
\end{equation}%
\begin{equation*}
\leq 2\dint\limits_{Q_{T}}\left( u_{t}-\beta \Delta u\right) ^{2}\varphi
_{\lambda ,\nu }dxdt+\nu \dint\limits_{Q_{T}}u^{2}\varphi _{\lambda ,\nu
}dxdt.
\end{equation*}%
Combining (\ref{2.11}) and (\ref{2.12}), we obtain (\ref{2.8}). $\square $

\section{Lipschitz Stability and Uniqueness}

\label{sec:4}

\textbf{Theorem 4.1.} \emph{Let }$M_{1},M_{2},M_{3},M_{4}>0$\emph{\ be
certain numbers. Assume that in (\ref{1.1}) the function }$F=F\left(
x,t,y,z\right) :\overline{Q}_{T}\times \mathbb{R}^{2}\rightarrow \mathbb{R}$%
\emph{\ be bounded in any bounded subset of the set }$\overline{Q}_{T}\times 
\mathbb{R}^{2}$\emph{\ and such that there exist derivatives }$%
F_{y},F_{z}\in C\left( \overline{Q}_{T}\times \mathbb{R}^{2}\right) $
satisfying 
\begin{equation}
\max \left( \sup_{Q_{T}\times \mathbb{R}^{2}}\left\vert F_{y}\left(
x,t,y,z\right) \right\vert ,\sup_{Q_{T}\times \mathbb{R}^{2}}\left\vert
F_{y}\left( x,t,y,z\right) \right\vert \right) \leq M_{1}.  \label{3}
\end{equation}%
\emph{Let the function }$K\left( x,y\right) $\emph{\ be bounded in }$%
\overline{\Omega }\times \overline{\Omega }$, \emph{the function }$\varkappa
\in C^{1}\left( \overline{\Omega }\right) $ \emph{and}%
\begin{equation}
\sup_{\Omega \times \Omega }\left\vert K\left( x,y\right) \right\vert
,\left\Vert \varkappa \right\Vert _{C^{1}\left( \overline{\Omega }\right)
}\leq M_{2}.  \label{3.100}
\end{equation}%
\emph{Consider the sets of functions }$B_{3}\left( M_{3}\right) ,B_{4}\left(
M_{4}\right) $\emph{\ defined as}%
\begin{equation}
B_{3}\left( M_{3}\right) =\left\{ u\in H_{0}^{2}\left( Q_{T}\right)
:\sup_{Q_{T}}\left\vert u\right\vert ,\sup_{Q_{T}}\left\vert \nabla
u\right\vert ,\sup_{Q_{T}}\left\vert \Delta u\right\vert \leq M_{3}\right\} ,
\label{3.2}
\end{equation}%
\begin{equation}
B_{4}\left( M_{4}\right) =\left\{ u\in H_{0}^{2}\left( Q_{T}\right)
:\sup_{Q_{T}}\left\vert u\right\vert ,\sup_{Q_{T}}\left\vert \nabla
u\right\vert \leq M_{4}\right\} ,  \label{3.02}
\end{equation}%
\emph{where the subspace }$H_{0}^{2}\left( Q_{T}\right) $ \emph{of the space 
}$H^{2}\left( Q_{T}\right) $ \emph{is the one defined in (\ref{2.00}).} 
\emph{Let}%
\begin{equation}
M=\max \left( M_{1},M_{2},M_{3},M_{4}\right) .  \label{3.002}
\end{equation}%
\emph{Assume that two pairs of functions }%
\begin{equation}
\left( v_{1},m_{1}\right) ,\left( v_{2},m_{2}\right) \in B_{3}\left(
M_{3}\right) \times B_{4}\left( M_{4}\right)  \label{3.0002}
\end{equation}%
\emph{\ satisfy equations (\ref{1.1}), zero Neumann boundary conditions (\ref%
{1.2}) as well as the following initial and terminal conditions:}%
\begin{equation}
v_{1}\left( x,T\right) =v_{T}^{\left( 1\right) }\left( x\right) ,\text{ }%
v_{2}\left( x,T\right) =v_{T}^{\left( 2\right) }\left( x\right) ,\text{ }%
x\in \Omega ,  \label{3.4}
\end{equation}%
\begin{equation}
m_{1}\left( x,T\right) =m_{T}^{\left( 1\right) }\left( x\right) ,\text{ }%
m_{2}\left( x,T\right) =m_{T}^{\left( 2\right) }\left( x\right) ,\text{ }%
x\in \Omega ,  \label{3.5}
\end{equation}%
\begin{equation}
m_{1}\left( x,0\right) =m_{0}^{\left( 1\right) }\left( x\right) ,m_{2}\left(
x,0\right) =m_{0}^{\left( 2\right) }\left( x\right) ,\text{ }x\in \Omega .
\label{3.6}
\end{equation}%
\emph{\ Then there exists a number }$C_{1}=C_{1}\left( \beta ,M,T\right) >0$ 
\emph{depending only on listed parameters such that the following two
Lipschitz stability estimates are valid:}%
\begin{equation}
\left. 
\begin{array}{c}
\left\Vert \partial _{t}v_{1}-\partial _{t}v_{2}\right\Vert _{L_{2}\left(
Q_{T}\right) }+\left\Vert \Delta v_{1}-\Delta v_{2}\right\Vert _{L_{2}\left(
Q_{T}\right) }+\left\Vert v_{1}-v_{2}\right\Vert _{H^{1,0}\left(
Q_{T}\right) }\leq \\ 
\leq C_{1}\left( \left\Vert v_{T}^{\left( 1\right) }-v_{T}^{\left( 2\right)
}\right\Vert _{H^{1}\left( \Omega \right) }+\left\Vert m_{T}^{\left(
1\right) }-m_{T}^{\left( 2\right) }\right\Vert _{L_{2}\left( \Omega \right)
}\right) + \\ 
+C_{1}\left\Vert m_{0}^{\left( 1\right) }-m_{0}^{\left( 2\right)
}\right\Vert _{H^{1}\left( \Omega \right) },%
\end{array}%
\right.  \label{3.7}
\end{equation}%
\begin{equation}
\left. 
\begin{array}{c}
\left\Vert m_{1}-m_{2}\right\Vert _{H^{1,0}\left( Q_{T}\right) }\leq
C_{1}\left( \left\Vert v_{T}^{\left( 1\right) }-v_{T}^{\left( 2\right)
}\right\Vert _{H^{1}\left( \Omega \right) }+\left\Vert m_{T}^{\left(
1\right) }-m_{T}^{\left( 2\right) }\right\Vert _{L_{2}\left( \Omega \right)
}\right) + \\ 
+C_{1}\left\Vert m_{0}^{\left( 1\right) }-m_{0}^{\left( 2\right)
}\right\Vert _{H^{1}\left( \Omega \right) }.%
\end{array}%
\right.  \label{3.8}
\end{equation}%
\emph{In particular, if the domain }$\Omega $ \emph{is a rectangular prism
as in (\ref{4.1}), then estimate (\ref{3.7}) is strengthened as}%
\begin{equation}
\left. 
\begin{array}{c}
\left\Vert v_{1}-v_{2}\right\Vert _{H^{2,1}\left( Q_{T}\right) }\leq
C_{1}\left( \left\Vert v_{T}^{\left( 1\right) }-v_{T}^{\left( 2\right)
}\right\Vert _{H^{1}\left( \Omega \right) }+\left\Vert m_{T}^{\left(
1\right) }-m_{T}^{\left( 2\right) }\right\Vert _{L_{2}\left( \Omega \right)
}\right) + \\ 
+C_{1}\left\Vert m_{0}^{\left( 1\right) }-m_{0}^{\left( 2\right)
}\right\Vert _{H^{1}\left( \Omega \right) }.%
\end{array}%
\right.  \label{3.70}
\end{equation}%
\emph{Next, if in (\ref{3.4})-(\ref{3.6})} 
\begin{equation}
v_{T}^{\left( 1\right) }\left( x\right) \equiv v_{T}^{\left( 2\right)
}\left( x\right) ,m_{0}^{\left( 1\right) }\left( x\right) \equiv
m_{0}^{\left( 2\right) }\left( x\right) ,m_{T}^{\left( 1\right) }\left(
x\right) \equiv m_{T}^{\left( 2\right) }\left( x\right) ,\text{ }x\in \Omega
,  \label{3.9}
\end{equation}%
\emph{then }$v_{1}\left( x,t\right) \equiv v_{2}\left( x,t\right) $\emph{\
and }$m_{1}\left( x,t\right) \equiv m_{2}\left( x,t\right) $ \emph{in }$%
Q_{T},$ \emph{which means that problem (\ref{1.1})-(\ref{1.3}) has at most
one solution.}

\textbf{Remark 4.1: }\emph{Our requirement that two pairs }$\left(
v_{1},m_{1}\right) ,\left( v_{2},m_{2}\right) $ \emph{belong to an a priori
chosen bounded set }$B_{3}\left( M_{3}\right) \times B_{4}\left(
M_{4}\right) $\emph{\ is a typical one in the theory of ill-posed problems.
Indeed, the solution of such a problem is typically thought for in an a
priori known bounded subset of a Banach space, rather than on the entire
Banach space, see, e.g. \cite{BK,LRS,T}. }

\textbf{Proof}. Note that (\ref{2.000}) implies that norms of functions in $%
H^{1}\left( \Omega \right) $ and $L_{2}\left( \Omega \right) $ in the right
hand sides of (\ref{3.7})-(\ref{3.70}) make sense. In this proof $%
C_{1}=C_{1}\left( \beta ,M,T\right) >0$ denotes different numbers depending
only on listed parameters. For the sake of definiteness, we set below the
parameter $a=2$ in (\ref{2.0}). Hence, below%
\begin{equation}
\varphi _{\lambda ,\nu }\left( t\right) =e^{2\lambda \left( t+2\right) ^{\nu
}},t\in \left( 0,T\right) .  \label{3.24}
\end{equation}

Denote 
\begin{equation}
\widetilde{v}\left( x,t\right) =v_{1}\left( x,t\right) -v_{2}\left(
x,t\right) ,\widetilde{m}\left( x,t\right) =m_{1}\left( x,t\right)
-m_{2}\left( x,t\right) ,\left( x,t\right) \in Q_{T}.  \label{3.10}
\end{equation}%
\begin{equation}
\widetilde{v}_{T}\left( x\right) =v_{T}^{\left( 1\right) }\left( x\right)
-v_{T}^{\left( 2\right) }\left( x\right) ,\text{ }\widetilde{m}_{T}\left(
x\right) =m_{T}^{\left( 1\right) }\left( x\right) -m_{T}^{\left( 2\right)
}\left( x\right) ,\text{ }x\in \Omega ,  \label{3.11}
\end{equation}%
\begin{equation}
\widetilde{m}_{0}\left( x\right) =m_{0}^{\left( 1\right) }\left( x\right)
-m_{0}^{\left( 2\right) }\left( x\right) ,\text{ }x\in \Omega .  \label{3.12}
\end{equation}

Subtract equations (\ref{1.1}) for the pair $\left( v_{2},m_{2}\right) $
from equations (\ref{1.1}) for the pair $\left( v_{1},m_{1}\right) .$ By (%
\ref{3.10}) 
\begin{equation}
m_{1}\nabla v_{1}-m_{2}\nabla v_{2}=\left( m_{1}\nabla v_{1}-m_{1}\nabla
v_{2}\right) +\left( m_{1}\nabla v_{2}-m_{2}\nabla v_{2}\right) =
\label{3.14}
\end{equation}%
\begin{equation*}
=m_{1}\nabla \widetilde{v}+\widetilde{m}\nabla v_{2}.
\end{equation*}%
Applying the multidimensional analog of Taylor formula \cite{V}, using (\ref%
{3.10}) and the first inequality (\ref{3.100}), we obtain%
\begin{equation}
\left. 
\begin{array}{c}
F\left( x,t,\dint\limits_{\Omega }K\left( x,y\right) m_{1}\left( y,t\right)
dy,m_{1}\left( x,t\right) \right) - \\ 
-F\left( x,t,\dint\limits_{\Omega }K\left( x,y\right) m_{2}\left( y,t\right)
dy,m_{2}\left( x,t\right) \right) = \\ 
=G_{1}\left( x,t\right) \dint\limits_{\Omega }K\left( x,y\right) \widetilde{m%
}\left( y,t\right) dy+G_{2}\left( x,t\right) \widetilde{m}\left( x,t\right) ,%
\end{array}%
\right.  \label{3.15}
\end{equation}%
where functions $G_{1}\left( x,t\right) $,$G_{2}\left( x,t\right) \in
C\left( \overline{Q}_{T}\right) $ and by (\ref{3}) 
\begin{equation}
\left\vert G_{1}\left( x,t\right) \right\vert ,\left\vert G_{2}\left(
x,t\right) \right\vert \leq M_{1},\text{ }\left( x,t\right) \in \overline{Q}%
_{T}.  \label{3.150}
\end{equation}%
Thus, using (\ref{3.100})-(\ref{3.0002}) and (\ref{3.10})-(\ref{3.150}), we
obtain the following two inequalities:%
\begin{equation}
\left\vert \widetilde{v}_{t}+\beta \Delta \widetilde{v}\right\vert \left(
x,t\right) \leq C_{1}\left( \left\vert \nabla \widetilde{v}\right\vert
+\left\vert \widetilde{m}\right\vert +\dint\limits_{\Omega }\left\vert 
\widetilde{m}\left( y,t\right) \right\vert dy\right) \left( x,t\right) ,%
\text{ }\left( x,t\right) \in Q_{T},  \label{3.17}
\end{equation}%
\begin{equation}
\left\vert \widetilde{m}_{t}-\beta \Delta \widetilde{m}\right\vert \left(
x,t\right) \leq C_{1}\left( \left\vert \nabla \widetilde{m}\right\vert
+\left\vert \widetilde{m}\right\vert +\left\vert \nabla \widetilde{v}%
\right\vert +\left\vert \Delta \widetilde{v}\right\vert \right) \left(
x,t\right) ,\text{ }\left( x,t\right) \in Q_{T}.  \label{3.18}
\end{equation}%
The boundary conditions for this system of two inequalities follow from (\ref%
{1.2})%
\begin{equation}
\left. 
\begin{array}{c}
\partial _{n}\widetilde{v}\mid _{S_{T}}=0, \\ 
\partial _{n}\widetilde{m}\mid _{S_{T}}=0.%
\end{array}%
\right.  \label{3.19}
\end{equation}%
Finally, initial and terminal conditions for system (\ref{3.17}), (\ref{3.18}%
) follow from (\ref{3.4})-(\ref{3.6}), (\ref{3.10}) and (\ref{3.11}), 
\begin{equation}
\widetilde{m}\left( x,0\right) =\widetilde{m}_{0}\left( x\right) ,
\label{3.20}
\end{equation}%
\begin{equation}
\widetilde{v}\left( x,T\right) =\widetilde{v}_{T}\left( x\right) ,
\label{3.21}
\end{equation}%
\begin{equation}
\widetilde{m}\left( x,T\right) =\widetilde{m}_{T}\left( x\right) .
\label{3.210}
\end{equation}

Square both sides of each of inequalities (\ref{3.17}), (\ref{3.18}), use
Cauchy-Schwarz inequality, multiply both sides of resulting two inequalities
by the Carleman Weight Function $\varphi _{\lambda ,\nu }\left( t\right) $
defined in (\ref{3.24}) and integrate over $Q_{T}.$ We obtain two integral
inequalities. The first one is:%
\begin{equation}
\dint\limits_{Q_{T}}\left( \widetilde{v}_{t}+\beta \Delta \widetilde{v}%
\right) ^{2}\varphi _{\lambda ,\nu }dxdt\leq C_{1}\dint\limits_{Q_{T}}\left(
\left( \nabla \widetilde{v}\right) ^{2}+\widetilde{m}^{2}+\dint\limits_{%
\Omega }\widetilde{m}^{2}\left( y,t\right) dy\right) \varphi _{\lambda ,\nu
}dxdt.  \label{3.22}
\end{equation}%
The second inequality is:%
\begin{equation}
\left. 
\begin{array}{c}
\dint\limits_{Q_{T}}\left( \widetilde{m}_{t}-\beta \Delta \widetilde{m}%
\right) ^{2}\varphi _{\lambda ,\nu }dxdt\leq \\ 
\leq C_{1}\dint\limits_{Q_{T}}\left( \left( \Delta \widetilde{v}\right)
^{2}+\left( \nabla \widetilde{v}\right) ^{2}+\left( \nabla \widetilde{m}%
\right) ^{2}+\widetilde{m}^{2}\right) \varphi _{\lambda ,\nu }dxdt.%
\end{array}%
\right.  \label{3.23}
\end{equation}

An element, which is unusual when working with Carleman estimates, is the
presence of the term $\left( \Delta \widetilde{v}\right) ^{2}$ in the second
line of (\ref{3.23}) since this is not a term with lower order derivatives.
Hence, we should somehow dominate this term when combining (\ref{3.23}) with
(\ref{3.22}), and this is done below.

It follows from (\ref{2.00}) and (\ref{3.19}) that we can apply Carleman
estimates of Theorems 3.1 and 3.2 to the left hand sides of (\ref{3.22}) and
(\ref{3.23}) respectively. Applying Carleman estimate (\ref{2.1}) of Theorem
3.1 with the Carleman Weight Function (\ref{3.24}) to the left hand side of (%
\ref{3.22}) and using (\ref{3.21}), we obtain%
\begin{equation}
\left. 
\begin{array}{c}
\dint\limits_{Q_{T}}\widetilde{v}_{t}^{2}\varphi _{\lambda ,\nu
}dxdt+\dint\limits_{Q_{T}}\left( \Delta \widetilde{v}\right) ^{2}\varphi
_{\lambda ,\nu }dxdt+ \\ 
+\lambda \nu \dint\limits_{Q_{T}}\left( \nabla \widetilde{v}\right)
^{2}\varphi _{\lambda ,\nu }dxdt+\lambda ^{2}\nu ^{2}\dint\limits_{Q_{T}}%
\widetilde{v}^{2}\varphi _{\lambda ,\nu }dxdt\leq \\ 
\leq C_{1}\dint\limits_{Q_{T}}\left( \left( \nabla \widetilde{v}\right) ^{2}+%
\widetilde{m}^{2}+\dint\limits_{\Omega }\widetilde{m}^{2}\left( y,t\right)
dy\right) \varphi _{\lambda ,\nu }dxdt+ \\ 
+C_{1}e^{3\lambda \left( T+2\right) ^{\nu }}\left\Vert \widetilde{v}%
_{T}\right\Vert _{H^{1}\left( \Omega \right) }^{2},\text{ }\forall \lambda
\geq 1,\forall \nu >2.%
\end{array}%
\right.  \label{3.25}
\end{equation}

Let $\nu _{0}>2$ be the number of Theorem 3.2. Assuming that $\nu \geq \nu
_{0}$ in (\ref{3.24}), we now apply Carleman estimate (\ref{2.8}) of Theorem
3.2 to the left hand side of (\ref{3.23}). Taking into account (\ref{3.20})
and (\ref{3.210}), we obtain%
\begin{equation}
\left. 
\begin{array}{c}
\sqrt{\nu }\dint\limits_{Q_{T}}\left( \nabla \widetilde{m}\right)
^{2}\varphi _{\lambda ,\nu }dxdt+\lambda \nu ^{2}\dint\limits_{Q_{T}}%
\widetilde{m}^{2}\varphi _{\lambda ,\nu }dxdt\leq \\ 
\leq C_{1}\dint\limits_{Q_{T}}\left( \left( \Delta \widetilde{v}\right)
^{2}+\left( \nabla \widetilde{v}\right) ^{2}+\left( \nabla \widetilde{m}%
\right) ^{2}+\widetilde{m}^{2}\right) \varphi _{\lambda ,\nu }dxdt+ \\ 
+C_{1}\lambda \nu \left( T+2\right) ^{\nu -1}e^{2\lambda \left( T+2\right)
^{\nu }}\left\Vert \widetilde{m}_{T}\right\Vert _{L_{2}\left( \Omega \right)
}^{2}+ \\ 
+C_{1}\sqrt{\nu }e^{2^{\nu +1}\lambda }\left\Vert \widetilde{m}%
_{0}\right\Vert _{H^{1}\left( \Omega \right) }^{2},\text{ }\forall \lambda
\geq 1,\forall \nu \geq \nu _{0}>2.%
\end{array}%
\right.  \label{3.26}
\end{equation}%
We now choose a number $\nu _{1}=\nu _{1}\left( \beta ,M,T\right) \geq \nu
_{0}$ depending only on listed parameters such that $\sqrt{\nu _{1}}\geq
2C_{1}.$ Next, choose the number $\lambda _{1}=\lambda _{1}\left( \beta
,M,T\right) \geq 1$ depending only on listed parameters such that $\lambda
_{1}\nu _{1}^{2}\geq 2C_{1}.$ Then (\ref{3.26}) implies%
\begin{equation}
\left. 
\begin{array}{c}
\dint\limits_{Q_{T}}\left( \nabla \widetilde{m}\right) ^{2}\varphi _{\lambda
,\nu _{1}}dxdt+\lambda \dint\limits_{Q_{T}}\widetilde{m}^{2}\varphi
_{\lambda ,\nu _{1}}dxdt\leq \\ 
\leq C_{1}\dint\limits_{Q_{T}}\left( \left( \Delta \widetilde{v}\right)
^{2}+\left( \nabla \widetilde{v}\right) ^{2}\right) \varphi _{\lambda ,\nu
_{1}}dxdt+ \\ 
+C_{1}e^{3\lambda \left( T+2\right) ^{\nu _{1}}}\left\Vert \widetilde{m}%
_{T}\right\Vert _{L_{2}\left( \Omega \right) }^{2}+C_{1}e^{2^{\nu
_{1}+1}\lambda }\left\Vert \widetilde{m}_{0}\right\Vert _{H^{1}\left( \Omega
\right) }^{2},\text{ }\forall \lambda \geq \lambda _{1}.%
\end{array}%
\right.  \label{3.27}
\end{equation}%
In particular, (\ref{3.27}) implies%
\begin{equation}
\left. 
\begin{array}{c}
\dint\limits_{Q_{T}}\widetilde{m}^{2}\varphi _{\lambda ,\nu
_{1}}^{2}dxdt\leq C_{1}\lambda ^{-1}\dint\limits_{Q_{T}}\left( \left( \Delta 
\widetilde{v}\right) ^{2}+\left( \nabla \widetilde{v}\right) ^{2}\right)
\varphi _{\lambda ,\nu _{1}}dxdt+ \\ 
+C_{1}e^{3\lambda \left( T+2\right) ^{\nu _{1}}}\left\Vert \widetilde{m}%
_{T}\right\Vert _{L_{2}\left( \Omega \right) }^{2}+C_{1}e^{2^{\nu
_{1}+2}\lambda }\left\Vert \widetilde{m}_{0}\right\Vert _{H^{1}\left( \Omega
\right) }^{2},\text{ }\forall \lambda \geq \lambda _{1}.%
\end{array}%
\right.  \label{3.28}
\end{equation}

The presence of the term $\left( \Delta \widetilde{v}\right) ^{2}/\lambda $
in (\ref{3.28}) with the large parameter $\lambda $ allows us to arrange the
above mentioned domination, which we need, see the paragraph below (\ref%
{3.23}).

Substituting (\ref{3.28}) in (\ref{3.25}) at $\nu =\nu _{1}>2,$ we obtain%
\begin{equation}
\left. 
\begin{array}{c}
\dint\limits_{Q_{T}}\widetilde{v}_{t}^{2}\varphi _{\lambda ,\nu
_{1}}dxdt+\dint\limits_{Q_{T}}\left( \Delta \widetilde{v}\right) ^{2}\varphi
_{\lambda ,\nu _{1}}dxdt+ \\ 
+\lambda \dint\limits_{Q_{T}}\left( \nabla \widetilde{v}\right) ^{2}\varphi
_{\lambda ,\nu _{1}}dxdt+\lambda ^{2}\dint\limits_{Q_{T}}\widetilde{v}%
^{2}\varphi _{\lambda ,\nu _{1}}dxdt\leq \\ 
\leq C_{1}\lambda ^{-1}\dint\limits_{Q_{T}}\left( \Delta \widetilde{v}%
\right) ^{2}\varphi _{\lambda ,\nu _{1}}dxdt+C_{1}\dint\limits_{Q_{T}}\left(
\nabla \widetilde{v}\right) ^{2}\varphi _{\lambda ,\nu _{1}}dxdt+ \\ 
+C_{1}e^{3\lambda \left( T+2\right) ^{\nu }}\left( \left\Vert \widetilde{v}%
_{T}\right\Vert _{H^{1}\left( \Omega \right) }^{2}+\left\Vert \widetilde{m}%
_{T}\right\Vert _{L_{2}\left( \Omega \right) }^{2}\right) \\ 
+C_{1}e^{2^{\nu _{1}+1}\lambda }\left\Vert \widetilde{m}_{0}\right\Vert
_{H^{1}\left( \Omega \right) }^{2},\text{ }\forall \lambda \geq \lambda _{1}.%
\end{array}%
\right.  \label{3.29}
\end{equation}%
Choose $\lambda _{2}=\lambda _{2}\left( \beta ,M,T\right) \geq \lambda
_{1}\geq 1$ depending only on listed parameters such that $2C_{2}/\lambda
_{2}\leq 1.$ Then (\ref{3.29}) implies%
\begin{equation}
\left. 
\begin{array}{c}
\dint\limits_{Q_{T}}\widetilde{v}_{t}^{2}\varphi _{\lambda ,\nu
_{1}}dxdt+\dint\limits_{Q_{T}}\left( \Delta \widetilde{v}\right) ^{2}\varphi
_{\lambda ,\nu _{1}}dxdt+ \\ 
+\dint\limits_{Q_{T}}\left( \nabla \widetilde{v}\right) ^{2}\varphi
_{\lambda ,\nu _{1}}dxdt+\dint\limits_{Q_{T}}\widetilde{v}^{2}\varphi
_{\lambda ,\nu _{1}}dxdt\leq \\ 
\leq C_{1}e^{3\lambda \left( T+2\right) ^{\nu }}\left( \left\Vert \widetilde{%
v}_{T}\right\Vert _{H^{1}\left( \Omega \right) }^{2}+\left\Vert \widetilde{m}%
_{T}\right\Vert _{L_{2}\left( \Omega \right) }^{2}\right) +C_{1}e^{2^{\nu
_{1}+1}\lambda }\left\Vert \widetilde{m}_{0}\right\Vert _{H^{1}\left( \Omega
\right) }^{2}, \\ 
\text{ }\forall \lambda \geq \lambda _{2}.%
\end{array}%
\right.  \label{3.30}
\end{equation}%
Combining (\ref{3.30}) with (\ref{3.27}), we obtain%
\begin{equation}
\left. 
\begin{array}{c}
\dint\limits_{Q_{T}}\left[ \left( \nabla \widetilde{m}\right) ^{2}+%
\widetilde{m}^{2}\right] \varphi _{\lambda ,\nu _{1}}dxdt\leq \\ 
\leq C_{1}e^{3\lambda \left( T+2\right) ^{\nu _{1}}}\left( \left\Vert 
\widetilde{v}_{T}\right\Vert _{H^{1}\left( \Omega \right) }^{2}+\left\Vert 
\widetilde{m}_{T}\right\Vert _{L_{2}\left( \Omega \right) }^{2}\right) + \\ 
+C_{1}e^{2^{\nu _{1}+1}\lambda }\left\Vert \widetilde{m}_{0}\right\Vert
_{H^{1}\left( \Omega \right) }^{2},\text{ }\forall \lambda \geq \lambda _{2}.%
\end{array}%
\right.  \label{3.31}
\end{equation}%
By (\ref{3.24}) 
\begin{equation*}
\min_{\left[ 0,T\right] }\varphi _{\lambda ,\nu _{1}}=\exp \left( 2^{\nu
_{1}+1}\lambda \right) .
\end{equation*}%
Hence, replacing in the left hand sides of (\ref{3.30}) and (\ref{3.31}) $%
\varphi _{\lambda ,\nu _{1}}$ with $e^{2^{\nu _{1}+1}\lambda },$ we
strengthen these inequalities. Hence, (\ref{3.30}) and (\ref{3.31}) lead to%
\begin{equation}
\left. 
\begin{array}{c}
\left\Vert \widetilde{v}_{t}\right\Vert _{L_{2}\left( Q_{T}\right)
}^{2}+\left\Vert \Delta \widetilde{v}\right\Vert _{L_{2}\left( Q_{T}\right)
}^{2}+\left\Vert \widetilde{v}\right\Vert _{H^{1,0}\left( Q_{T}\right)
}^{2}\leq \\ 
\leq C_{1}\left[ e^{3\lambda \left( T+2\right) ^{\nu _{1}}}\left( \left\Vert 
\widetilde{v}_{T}\right\Vert _{H^{1}\left( \Omega \right) }^{2}+\left\Vert 
\widetilde{m}_{T}\right\Vert _{L_{2}\left( \Omega \right) }^{2}\right)
+\left\Vert \widetilde{m}_{0}\right\Vert _{H^{1}\left( \Omega \right) }^{2}%
\right] , \\ 
\forall \lambda \geq \lambda _{2},%
\end{array}%
\right.  \label{3.32}
\end{equation}%
\begin{equation}
\left. 
\begin{array}{c}
\left\Vert \widetilde{m}\right\Vert _{H^{1,0}\left( Q_{T}\right) }^{2}\leq
\\ 
\leq C_{1}\left[ e^{3\lambda \left( T+2\right) ^{\nu _{1}}}\left( \left\Vert 
\widetilde{v}_{T}\right\Vert _{H^{1}\left( \Omega \right) }^{2}+\left\Vert 
\widetilde{m}_{T}\right\Vert _{L_{2}\left( \Omega \right) }^{2}\right)
+\left\Vert \widetilde{m}_{0}\right\Vert _{H^{1}\left( \Omega \right) }^{2}%
\right] , \\ 
\forall \lambda \geq \lambda _{2}.\text{ }%
\end{array}%
\right.  \label{3.320}
\end{equation}%
Hence, setting in (\ref{3.32}), (\ref{3.32}) $\lambda =\lambda _{2}=\lambda
_{2}\left( \beta ,M,T\right) \geq 1,$ we obtain%
\begin{equation}
\left. 
\begin{array}{c}
\left\Vert \widetilde{v}_{t}\right\Vert _{L_{2}\left( Q_{T}\right)
}+\left\Vert \Delta \widetilde{v}\right\Vert _{L_{2}\left( Q_{T}\right)
}+\left\Vert \widetilde{v}\right\Vert _{H^{1,0}\left( Q_{T}\right) }^{2}\leq
\\ 
\leq C_{1}\left( \left\Vert \widetilde{v}_{T}\right\Vert _{H^{1}\left(
\Omega \right) }+\left\Vert \widetilde{m}_{T}\right\Vert _{L_{2}\left(
\Omega \right) }+\left\Vert \widetilde{m}_{0}\right\Vert _{H^{1}\left(
\Omega \right) }\right) ,%
\end{array}%
\right.  \label{3.33}
\end{equation}%
\begin{equation}
\left\Vert \widetilde{m}\right\Vert _{H^{1,0}\left( Q_{T}\right) }^{2}\leq
C_{1}\left( \left\Vert \widetilde{v}_{T}\right\Vert _{H^{1}\left( \Omega
\right) }+\left\Vert \widetilde{m}_{T}\right\Vert _{L_{2}\left( \Omega
\right) }+\left\Vert \widetilde{m}_{0}\right\Vert _{H^{1}\left( \Omega
\right) }\right) .  \label{3.34}
\end{equation}%
It follows from (\ref{3.10})-(\ref{3.12}) that estimates (\ref{3.33}) and (%
\ref{3.34}) are equivalent with target estimates (\ref{3.7}) and (\ref{3.8})
respectively. If the domain $\Omega $ is a rectangular prism as in (\ref{4.1}%
), then estimate (\ref{3.70}) follows immediately from (\ref{3.7}) and Lemma
2.1. $\square $

\section{Appendix: A Brief Outline of the Derivation of MFGS (\protect\ref%
{1.1})}

\label{sec:5}

In this section, we briefly outline the derivation of MFGS (\ref{1.1}) from
the setting of the stochastic differential game with infinitely many similar
players in a bounded domain. We follow the approach of \cite[\S 3.1.2]%
{Carmona_Delarue_I} and \cite{Lasry_Lions_2006_I,Lasry_Lions_2006_II}. Since
a detailed derivation would be quite space consuming, we are rather brief
here and refer to these references for details.

Our plan is the following: we start with the optimal control problem for
stochastic differential equation with reflection; then we deduce the
Fokker-Planck equation for that SDE; finally combining first two results, we
derive MFGS (\ref{1.1}).

We consider a stochastic differential equation with reflection (SDER) on a
bounded domain. For simplicity, we assume the constant volatility. Let $%
f\left( x,t\right) $ be a function defined on $Q_{T}$ with values in $%
\mathbb{R}^{n}$, $W_{t}$ be an $n-$dimensional Wiener process defined on a
filtered probability space $(\Omega ,\mathcal{F},\{\mathcal{F}_{t}\}_{t\in
\lbrack 0,T]},\mathbb{P})$. The coefficient $\beta >0$ indicates the
intensity of the white noise, while the function $f\left( x,t\right) $
describes the drift. The following definition is borrowed from \cite%
{Dupuis_Ishii} and \cite{Pilipenko}.

\begin{definition}
We say that a pairs of continuous $\{\mathcal{F}_{t}\}_{t\in \lbrack 0,T]}-$%
adapted processes $(X_{t},l(t))$ is a strong solution of the reflecting SDE 
\begin{equation}
dX_{t}=f(X_{t},t)dt+\sqrt{2\beta }dW_{t}+n(X_{t})dl(t)  \label{eq_itro:sder}
\end{equation}%
with the initial condition $X(0)=x_{0}$ if the following conditions hold true

\begin{itemize}
\item $X_{t}\in \overline{\Omega }$ $\mathbb{P}-$almost sure;

\item the function $l(t)$ is non-decreasing for $t\in \lbrack 0,T]$, $l(0)=0$
and 
\begin{equation*}
\dint\limits_{0}^{T}\Vert n(X_{t})\Vert dl(t)<\infty ,\ \
\dint\limits_{0}^{T}\mathbbm{1}_{\func{int}(\Omega )}(X_{t})dl(t)=0;
\end{equation*}

\item 
\begin{equation}
X_{t}=x_{0}+\dint\limits_{0}^{t}f(s,X_{s})ds+\sqrt{2\beta }%
W_{t}+\dint\limits_{0}^{t}n(X_{s})dl(s),  \label{equality:sder}
\end{equation}%
and all integrals in \eqref{equality:sder} are well-defined.
\end{itemize}
\end{definition}

Here $\mathbbm{1}_{A}$ stands for the indicator function for a set $A$,
i.e., $\mathbbm{1}_{A}(x)=1$ when $x\in A$ and $\mathbbm{1}_{A}(x)=0$ in the
opposite case. Note that if the function $f\left( x,t\right) $ is continuous
and Lipschitz continuous with respect to $x$, then there exits a unique
strong solution to \eqref{eq_itro:sder} \cite[Corollary 5.2]{Dupuis_Ishii}.

Let us turn now to the optimal control problem and consider the controlled
system 
\begin{equation}
dX_{t}=\varkappa (X_{t})\alpha (X_{t},t)dt+\sqrt{2\beta }dW_{t}+n(X_{t})dl(t)
\label{eq_itro:sder_control_fb}
\end{equation}%
Here $\alpha (x,t)$ is a control applied at $(x,t)$. The function $\varkappa
:\Omega \rightarrow \mathbb{R}$ is a characteristics of the medium. This
function characterizes the reaction of the controlled object to an action
applied at the point $x$. Let $t_{0}\in \lbrack 0,T]$ be an arbitrary
number, and let $x_{0}\in \mathbb{R}^{n}$ be an arbitrary point. It is
convenient to denote the solution of \eqref{eq_itro:sder_control_fb} with
the initial condition $X(t_{0})=x_{0}$ by $X_{t}^{x_{0},t_{0},\alpha }$. The
quality of the control is evaluated by a payoff functional 
\begin{equation}
J(x_{0},t_{0},\alpha )\triangleq \mathbb{E}\Bigg[v_{T}(X_{T}^{x_{0},t_{0},%
\alpha })+\dint\limits_{t_{0}}^{T}\Bigg(-\frac{\alpha
^{2}(X_{t}^{x_{0},t_{0},\alpha },t)}{2}+G(X_{t}^{x_{0},t_{0},\alpha },t)%
\Bigg)dt\Bigg].  \label{eq_intro:payoff_fb}
\end{equation}%
In (\ref{eq_intro:payoff_fb}), one can interpret the term $\alpha
^{2}(X_{t}^{x_{0},t_{0},\alpha },t)/2$ as an instantaneous energy
consumption, while $G(X_{t}^{x_{0},t_{0},\alpha },t)$ evaluates the path.
This term plays the role of the potential energy. In the following, we
assume that the function $G\left( x,t\right) $ is continuous and, for each $%
t\in \left[ 0,T\right] $ it has at most linear growth with respect to~$%
\left\vert x\right\vert $.

The key notion in the control theory is the value function that is defined
by the rule: 
\begin{equation}
\func{Val}(x_{0},t_{0})\triangleq \sup_{\alpha \in \mathcal{A}}\Big\{%
J(x_{0},t_{0},\alpha )\}  \label{intro:Val}
\end{equation}%
Here we denoted by $\mathcal{A}$ the set of controls $\alpha
:Q_{T}\rightarrow \mathbb{R}^{n}$ which are measurable and Lipschitz
continuous with respect to $x$.

A strategy $\alpha \left( x,t\right) $ is optimal if it provides the maximum
in the right-hand side of \eqref{intro:Val} for each $(x_{0},t_{0})\in Q_{T}$%
.

The value function and the optimal strategy can be obtained through the
dynamic programming principle that reduces an optimal control problem to a
PDE \cite[\S IV.3]{Fleming_Soner}. For problem~%
\eqref{eq_itro:sder_control_fb},~\eqref{eq_intro:payoff_fb}, the Bellman
equation takes the form 
\begin{equation}
v_{t}(x,t)+\beta \Delta v(x,t)+\varkappa ^{2}(x)(\nabla
v(x,t))^{2}/2+G(x,t)=0  \label{eq:Bellman}
\end{equation}

\begin{theorem}
\label{th:verification} Let $v(x,t)$ be the $C^{2,1}\left( \overline{Q}%
_{T}\right) -$solution of Bellman equation \eqref{eq:Bellman} with the
boundary condition 
\begin{equation}
\partial _{n}v(x,t)=0,\ \ (x,t)\in S_{T},\ \ t\in (0,T)
\label{cond:neumann_V}
\end{equation}%
and the terminal condition%
\begin{equation}
v(T,x)=v_{T}(x),\ \ x\in \Omega .  \label{cond:final_V}
\end{equation}%
Let the strategy $\hat{\alpha}(x,t)$ be defined by the rule: 
\begin{equation}
\hat{\alpha}(x,t)\triangleq \varkappa (x)\nabla v(x,t).  \label{2}
\end{equation}%
Then, $v(x_{0},t_{0})=\func{Val}(x_{0},t_{0})$ for each $(x_{0},t_{0})\in
Q_{T}$, and $\hat{\alpha}(x,t)$ is the unique optimal strategy.
\end{theorem}

The proof of this theorem can be obtained via standard verification
arguments (see, for example, \cite[Theorem IV.3.1, Remark IV.3.2]%
{Fleming_Soner}) and it uses the technique of \cite[Theorem 3.1.1]{Pilipenko}%
.


We now consider the dynamics given by \eqref{eq_intro:payoff_fb} assuming
that the function $\alpha \left( x,t\right) $ is Lipschitz continuous with
respect to $x$, and the initial state is distributed on $\Omega $ with a
probability determined by its density $m_{0}\left( x\right) $. In this case
(see \cite[Remark 3.1.2]{Pilipenko}), if $m:Q_{T}\rightarrow \mathbb{R}$ is
the $C^{2,1}\left( \overline{Q}_{T}\right) -$solution of the Fokker-Planck
equation 
\begin{equation}
m_{t}(x,t)-\beta \Delta m(x,t)+\nabla \cdot (\varkappa (x)\alpha
(x,t)m(x,t))=0  \label{eq:FP}
\end{equation}%
with the boundary condition 
\begin{equation}
\beta \partial _{n}m(x,t)-\varkappa (x)\langle \alpha (x,t),n(x)\rangle
m(x,t)=0,\ \ (x,t)\in S_{T},  \label{cond:FP}
\end{equation}%
and the initial condition 
\begin{equation}
m(x,0)=m_{0}\left( x\right) ,\ \ x\in \Omega ,  \label{cond:initial_FP}
\end{equation}%
then, for each $t\in \lbrack 0,T]$, the function $x\mapsto m(x,t)$ is the
density for the distribution $X_{t}$. In (\ref{cond:FP}) and below $%
\left\langle ,\right\rangle $ denotes the scalar product in $\mathbb{R}^{n}.$

The mean field game system arises in the study of a Nash equilibrium for a
stochastic differential game with infinitely many identical players \cite%
{Lasry_Lions_2006_I}, \cite{Lasry_Lions_2006_II}. To derive MFGS (\ref{1.1}%
), we assume that each player chooses a control $\alpha \left( x,t\right) $
and tries to maximize the payoff functional 
\begin{equation}
\begin{split}
\mathbb{E}\Bigg[v_{T}(X_{T}& )+\dint\limits_{t_{0}}^{T}\Bigg(-\frac{\alpha
^{2}(X_{t},t)}{2}+ \\
& +F\Big(X_{t},t,\dint\limits_{\Omega }K(X_{t},y)m(y,t),m(X_{t},t)\Big)\Bigg)%
dt\Bigg],
\end{split}
\label{eq_intro:payoff_mfg}
\end{equation}%
subject to dynamics \eqref{eq_itro:sder_control_fb} and the initial
distribution of players $m_{0}$. Payoff \eqref{eq_intro:payoff_mfg} is a
generalization of functional \eqref{eq_intro:payoff_fb}. Here $m(x,t)$ is
the density of the distribution of all players. Since the players are
identical, then the desired MFG equilibrium implies that $m(x,t)$ is also
the distribution of $X_{t}$ for each player. We now replace the function $G$
in \eqref{eq_intro:payoff_fb} with the function $F$, which can be regarded
as a potential energy. However, in contrast with \eqref{eq_intro:payoff_fb},
it depends now not only on the current state of the player, but also on the
density of the distribution of all players.

\textbf{Remark 5.1. }\emph{Note that one of the advantages of our technique
is that it allows to combine the convolution operator of the long range
interaction with operator of the short range, or local, interaction.}

In particular, the considered class of interaction terms includes the
non-local dependence given by a convolution operator, i.e. the integral
operator with its kernel $K(x,y)$. For example, one can consider a model of
the long-range interaction with 
\begin{equation*}
F\Big(x,t,\dint\limits_{\Omega }K(x,y)m(y,t)dy,m(x,t)\Big)%
=U(x,t)+\dint\limits_{\Omega }K(x,y)m(y,t)dy.
\end{equation*}%
In this case, $U:Q_{T}\rightarrow \mathbb{R}$ indicates the action on a
player of an external force, while the integral%
\begin{equation}
\dint\limits_{\Omega }K(x,y)m(y,t)dy  \label{1}
\end{equation}%
describes the inner forces. The function $K(x,y)$ is the action on the
player, who occupies the state $x,$ by the player, who occupies the state $y$%
. Thus, the integral (\ref{1}) is the average action on a player, who
occupies the state $x,$ by all other players.

If the radius of the interaction tends to zero, then we arrive at the model
of short-range interaction. A good limiting model in this case is 
\begin{equation*}
F\Big(x,t,\dint\limits_{\Omega }K(x,y)m(y,t)dy,m(x,t)\Big)=U(x,t)+k(x)m(x,t).
\end{equation*}%
Here $k:\Omega \rightarrow \mathbb{R}$ is a function characterizing this
short-range interaction.

Since the Nash equilibrium for a system of identical player is studied,
while the unilateral change of the strategy does not affect the distribution
of all player, we obtain that each player solves the optimal control problem %
\eqref{eq_itro:sder_control_fb}, \eqref{eq_intro:payoff_fb} for 
\begin{equation*}
G(x,t)=F\Big(x,t,\dint\limits_{\Omega }K(x,y)m(y,t)dy,m(x,t)\Big),
\end{equation*}%
the given function $m\left( x,t\right) $ and initial distribution $%
m_{0}\left( x\right) $. Due to Theorem \ref{th:verification}, the value
function solves the equation 
\begin{equation}
v_{t}+\beta \Delta v+\varkappa ^{2}(x)(\nabla v(x,t))^{2}/2+F\Big(%
x,t,\dint\limits_{\Omega }K(x,y)m(y,t)dy,m(x,t)\Big)=0
\label{eq:Bellman_mfg}
\end{equation}%
with conditions \eqref{cond:neumann_V}, \eqref{cond:final_V}, while the
optimal strategy is 
\begin{equation*}
\hat{\alpha}(x,t)=\varkappa (x)\nabla v(x,t).
\end{equation*}%
Plugging this into \eqref{eq:FP}, we arrive at the following Fokker-Planck
equation 
\begin{equation}
m_{t}(x,t)-\beta \Delta m(x,t)+\nabla \cdot (\varkappa ^{2}(x)\nabla
v(x,t)m(x,t))=0.  \label{eq:FP_mfg}
\end{equation}

Furthermore, \eqref{cond:FP} takes the form 
\begin{equation*}
\beta \partial _{n}m(x,t)-\varkappa ^{2}(x)\langle \nabla v(x,t),n(x)\rangle
m(x,t)=0,\ \ (x,t)\in S_{T}.
\end{equation*}%
The second term here equals to zero due to \eqref{cond:neumann_V}.
Therefore, we arrive at the Neumann boundary condition 
\begin{equation}
\partial _{n}m(x,t)=0,\ \ (x,t)\in S_{T}.  \label{cond:mfg_boundary}
\end{equation}

Equations \eqref{eq:Bellman_mfg}, \eqref{eq:FP_mfg} and conditions %
\eqref{cond:neumann_V}, \eqref{cond:final_V}, \eqref{cond:initial_FP}, %
\eqref{cond:mfg_boundary} form a MFG system on the bounded domain $\Omega $
in the case when the interaction term $F$ includes both the nonlocal
interaction given by the convolution (\ref{1}) and the local interaction
determined directly by the density $m(x,t)$. Moreover, if the solution of
MFGS (\ref{1.1}) is given, then the strategy $\hat{\alpha}\left( x,t\right) $
in (\ref{2}) provides the Nash equilibrium in the considered stochastic
differential game with infinitely many identical players.

\end{document}